\documentclass[11pt,a4paper,twoside]{article}
\usepackage{amsmath}
\usepackage{amsfonts}
\usepackage{changes}
\numberwithin{equation}{section}

\begin{document}

\noindent

{\bf
{\Large From Chern-Simons to Tomonaga-Luttinger \\ 

}
} 

\vspace{.5cm}
\hrule

\vspace{1cm}

\noindent

{\large\bf{Nicola Maggiore\footnote{\tt nicola.maggiore@ge.infn.it }
\\[1cm]}}

\setcounter{footnote}{0}

\noindent
{{}Dipartimento di Fisica, Universit\`a di Genova,\\
via Dodecaneso 33, I-16146, Genova, Italy\\
and\\
{} I.N.F.N. - Sezione di Genova\\
\vspace{1cm}

\noindent
{\tt Abstract~:}
A single-sided boundary is introduced in the three-dimensional Chern-Simons model. It is shown that only one boundary condition for the gauge fields is possible, which plays the twofold role of chirality condition and bosonization rule for the two-dimensional Weyl fermion describing the degrees of freedom of the edge states of the Fractional Quantum Hall Effect. It is derived the symmetry on the boundary which determines the effective two dimensional action, whose equation of motion coincides with the continuity equation of the Tomonaga-Luttinger theory. The role of Lorentz symmetry and of discrete symmetries on the boundary is also discussed.

\newpage

\section{Introduction}

The only way to give local observables to a Topological Quantum Field Theory (TQFT)  is to introduce a boundary, in absence of which the only observables are the global geometrical properties of the manifold on which the TQFT lives \cite{Birmingham:1991ty}. It is known since long time that quantizing a TQFT on a manifold with a boundary, one can recover all the states of the chiral algebra and its representations \cite{Moore:1989yh}. For non abelian Chern-Simons (CS) model, it has been shown that chiral currents satisfying a 
Ka\c{c}-Moody algebra live on the two-dimensional (2D) boundary of the 3D bulk manifold \cite{Blasi:1990pf, Emery:1991tf}. This fact is particularly relevant in condensed matter physics, since the 2D chiral currents on the boundary of the 3D CS model have been identified with the edge states in the Laughlin series of the Fractional Quantum Hall Effect (FQHE) \cite{Wen:1990se}. In \cite{Blasi:2008gt} a double-sided boundary for the CS theory has been introduced following Symanzik's approach \cite{Symanzik:1981wd}, which consists in imposing a separability condition on the propagators of the theory. These latter, once restricted on the boundary, are interpreted as the algebra formed by currents, which are shown to be chiral, thanks to certain boundary conditions. The problem of uniquely identifying a 2D effective boundary action and the corresponding degrees of freedom was left open. A 2D action was argued, and it was checked $a\ posteriori$ that indeed it represented the correct result \cite{Blasi:2008gt}.

In this paper we face the problem in a different way and with a different perspective, without imposing the Symanzik's separability condition on the propagators, rather focusing on the bulk action, on which a planar, single-sided boundary is introduced, as done in \cite{Amoretti:2014kba}. The boundary term in the action is realized coupling a $\delta(x)$ distribution to a functional depending on the 3D gauge fields by means of which the the CS action is built. 

The advantages of this approach are many. The most evident one is that it is not necessary to compute the propagators of the theory with boundary, which, in general, could be a very difficult task \cite{Blasi:2010gw}. More important, the approach we follow in this paper allows for a unique and unambiguous determination of the degrees of freedom which live on the boundary. In fact, we will be able to identify the symmetry which uniquely identifies the effective 2D boundary action. Moreover, we will be able to isolate the only possible boundary condition, which we will show to play the twofold role of bosonization and chirality condition for a 2D Weyl fermion living on the boundary. This result  rigorously clarifies the claim according to which on the boundary of the bosonic CS action reside the edge fermionic chiral states of the FQHE.

The paper is organized as follows. 

In Section 2 the total 3D CS bulk action is constructed. First of all, the planar boundary $x_2=0$ is introduced in the bulk 3D CS action by means of a Heaviside step function $\theta(x_2)$. Then, a gauge-fixing term is added, corresponding to the non-covariant axial choice $A_2(x)=0$. Moreover,  external sources are coupled, as usual, to the 3D gauge fields. Finally, the $\delta(x_2)$-boundary term is constructed, in terms of the two gauge fields surviving the axial gauge choice. The way in which the boundary term is introduced is one of the novelties contained in this paper. In fact, on the planar boundary $x_2=0$ not even the 2D Lorentz invariance has been invoked in order to write it, differently from what has been done previously \cite{Amoretti:2014iza}. This will turn out to be crucial, because only for a non-covariant boundary term it will be possible to get nontrivial results.

In Section 3 we derive the equations of motion for the whole action, from which we will get both the boundary conditions and the residual Ward identity, whose existence is due to the fact that the axial gauge-fixing does not completely fix the gauge \cite{Bassetto:1991ue}. In addition, in this Section a detailed study of the possible discrete symmetries of the total action is performed. The generic boundary term breaks any symmetry, and the conditions for invariance under each discrete symmetry are provided. Basically, there are two possible discrete symmetries, which can be traced back to parity and time reversal, with all the $caveat$ due to the euclidean choice of the flat spacetime. 

In Section 4 the boundary conditions found in the previous Section are solved, with and without imposing the discrete symmetries. The outcome is that the possible compatible boundary conditions fall into three categories.

In Section 5, the 2D algebra is derived from the residual Ward identity, and the boundary degree of freedom, which turns out to be a scalar field, is identified. Moreover, from the algebra the generalized canonical variables are identified, which lead to the effective 2D action compatible with the 2D symmetry and the boundary conditions found previously.

In Section 6 contact is made with condensed matter physics. In particular, the boundary condition  is interpreted as a bosonization condition for a 2D Weyl fermion, and the equation of motion derived from the 2D bosonic action is rewritten as the continuity equation for the Tomonaga-Luttinger theory \cite{Tomonaga:1950zz,Luttinger:1963zz,Haldane:1981zza}.

Finally, Section 7 summarizes our results.

\section{The action with boundary}

We are working in the flat three dimensional euclidean spacetime, described by the metric  $\eta_{\mu\nu}=\mbox{diag}(1,1,1)$. Our conventions concerning greek and latin indices are as follows: 
$\mu=0,1,2\ ;\ i=0,1$. The Levi-Civita completely antisymmetric tensor is $\epsilon_{\mu\nu\rho}$, with $\epsilon_{012}=1$ and $\epsilon_{2ij}\equiv\epsilon_{ij}$.

The CS action is given by
\begin{equation}
S_{CS}=\frac{k}{2}\int d^3x\ \epsilon_{\mu\nu\rho}A_\mu\partial_\nu A_\rho \ ,
\label{2.1}\end{equation}
where we choose to maintain the constant $k$ although it could be reabsorbed by a redefinition of the gauge field $A_\mu(x)$, in order to keep trace of the contribution of the bulk CS action in what follows.

We now introduce a boundary at $x_2=0$, which is implemented by means of  the Heaviside step function 
$\theta(x_2)$:
\begin{eqnarray}
S_{bulk}&=&
\frac{k}{2}\int d^3x\ \theta(x_2)\epsilon_{\mu\nu\rho}A_\mu\partial_\nu A_\rho\nonumber\\
&=& 
\frac{k}{2}\int d^3x\ \theta(x_2)
\left[A_0(\partial_1A_2-\partial_2A_1) - A_1(\partial_0A_2-\partial_2A_0)\right.\nonumber \\
&&\ \ \ \ \ \ \ \ \ \ \ \left. +A_2(\partial_0A_1-\partial_1A_0)\right].
\label{2.2}\end{eqnarray}

Under the infinitesimal gauge transformation 
\begin{equation}
\delta_{gauge} A_\mu=\partial_\mu\lambda(x),
\label{2.3}\end{equation}
where $\lambda(x)$ is a local gauge parameter, the CS action \eqref{2.1} is left invariant
\begin{equation}\delta_{gauge} S_{CS}=0.
\label{2.4}\end{equation}

The presence of the boundary breaks the gauge invariance of the CS action
\begin{equation}
\delta_{gauge} S_{bulk}=-\frac{k}{2}\int d^3x\ \delta(x_2)\epsilon_{ij}\lambda\partial_iA_j \ ,
\label{2.5}\end{equation}
where we used $\partial_\mu\theta(x_2)=\delta_{2\mu}\delta(x_2)$.

The total action we consider is 
\begin{equation}
S_{tot}=S_{bulk}+S_{gf}+S_J+S_{bd} \ ,
\label{2.6}\end{equation}
where $S_{bulk}$ is the bulk CS action \eqref{2.2} with boundary at $x_2=0$, $S_{gf}$ is the gauge fixing term
\begin{equation}
S_{gf}=\int d^3x\ \theta(x_2)bA_2 \ ,
\label{2.7}\end{equation}
where $b(x)$ is a Lagrange multiplier enforcing the gauge fixing choice $A_2=0$, $S_J$ is the source term 
\begin{equation}
S_{J}=\int d^3x\ \theta(x_2)J_iA_i \ ,
\label{2.8}\end{equation}
where $J_i$ are two external sources coupled to $A_i$, and, finally, $S_{bd}$ is the most general boundary action living on the boundary $x_2=0$
\begin{equation}
S_{bd}=\int d^3x\ \delta(x_2)\left(
\frac{a_1}{2}A_0^2+a_2A_0A_1+\frac{a_3}{2}A_1^2\right).
\label{2.9}\end{equation}
The boundary action $S_{bd}$ \eqref{2.9} depends on three constants $a_1$, $a_2$ and $a_3$. Notice that, unless $a_1=a_3$ and $a_2=0$, the boundary action $S_{bd}$ breaks Lorentz invariance, as already done by the gauge fixing term $S_{gf}$ \eqref{2.7} and by the presence of the boundary $x_2=0$ in the bulk action 
$S_{bulk}$ \eqref{2.2}, so that Lorentz invariance should not be a mandatory request. The Lorentz invariant  boundary action depends on one constant only:
\begin{equation}
S_{bd}=\int d^3x\ \delta(x_2)
\frac{a}{2}A_i^2 \ .
\label{2.10}\end{equation}
The role of a non-Lorentz invariant boundary term in the total action $S_{tot}$ \eqref{2.6} is crucial, as it will be clear in the next sections. As a remark concerning non Lorentz-invariant terms, like the ones contained in \eqref{2.9}, added to otherwise covariant actions, we mention the case of massive gravity. In that framework, indeed, it is relevant to find mass terms for the linearized graviton which lie outside the Fierz-Pauli paradigm \cite{Fierz:1939ix}. An important step towards that goal has been done in \cite{Rubakov:2004eb,Libanov:2005vu}, where Lorentz-violating mass terms have been added to the theory of linearized gravity. Once the path to alternative massive gravities has been opened, more general mass terms for massive gravity have been found in \cite{Blasi:2015lrg,Blasi:2017pkk}.

\section{Equations of motion, Ward identity and discrete symmetries}

For the total action $S_{tot}$ \eqref{2.6}, the equations of motion read
\begin{eqnarray}
\frac{\delta S_{tot}}{\delta A_0} &=& 
\theta(x_2)[k(\partial_1A_2-\partial_2A_1)+J_0] 
\nonumber \\
&&
+\delta(x_2)[a_1A_0+(a_2-k)A_1]=0 \label{3.1} \\
\frac{\delta S_{tot}}{\delta A_1} &=& 
\theta(x_2)[k(\partial_2A_0-\partial_0A_2)+J_1] 
\nonumber \\
&&
+\delta(x_2)[a_3A_1+(a_2+k)A_0]=0 \label{3.2}\\
\frac{\delta S_{tot}}{\delta A_2} &=& 
\theta(x_2)[k(\partial_0A_1-\partial_1A_0)+b]=0.  \label{3.3}
\end{eqnarray}
From the equations of motion, the following integrated Ward identity is easily derived
\begin{equation}
\int_0^\infty dx_2\ \partial_iJ_i=\left.-k\epsilon_{ij}\partial_iA_j\right|_{x_2=0}. 
\label{3.4}\end{equation}
Putting equal to zero the $\delta(x_2)$ boundary term in the equations of motion \eqref{3.1} and \eqref{3.2}, we get the boundary conditions on $x_2=0$
\begin{eqnarray}
a_1A_0+(a_2-k)A_1 &=& 0\label{3.5}\\
a_3A_1+(a_2+k)A_0 &=&0\label{3.6}.
\end{eqnarray}
The bulk action $S_{bulk}$ \eqref{2.2} displays two discrete symmetries: 

\begin{enumerate}
\item
$P$-symmetry (``parity''): $x_0\rightarrow x_0\ ;\ x_i\rightarrow -x_i$

\begin{equation}
PA_0=A_0\ \ ;\ \ PA_1=-A_1\ \ ;\ \ PA_2=-A_2
\label{3.7}\end{equation}

\item 
$R$-symmetry: $x_0\leftrightarrow x_1 \ ;\  x_2\rightarrow -x_2$

\begin{equation}
RA_0=-A_1\ ;\ RA_1=-A_0\ ;\  RA_2=A_2 \ .
\label{3.8}\end{equation}
\end{enumerate}

Imposing the discrete symmetries $P$ and $R$ on the boundary action $S_{bd}$ \eqref{2.9}, yields the following constraints on the parameters $a_i$:
\begin{center}
\begin{equation}
\begin{tabular}{|l|l|l|l|}
\hline
& $P$ &$R$ & \mbox{Lorentz}\\ \hline
$a_1$&$=a_1$ &$=a_1$&$=a_1$ \\ \hline
$a_2$&$=0$ &$=a_2$&$=0$ \\ \hline
$a_3$&$=a_3$ & $=a_1$&$=a_1$ \\ \hline
\end{tabular}
\label{3.9}
\end{equation}
\end{center}

where we added also the constraints deriving from the Lorentz symmetry, noticing that the latter can be seen as a combination of $P$ and $R$: 
\begin{equation}
PR=RP=\mbox{Lorentz}.
\label{3.10}\end{equation}

\section{Boundary conditions}

Let us now look for solutions of the boundary conditions \eqref{3.5} and \eqref{3.6}, with and without imposing the discrete symmetries \eqref{3.7} and \eqref{3.8}.
\begin{enumerate}
\item Lorentz: $a_1=a_3\ \ ;\ \ a_2=0$ 

no solution

\item P-symmetry $a_1=a_1\ \ ;\ \ a_2=0\ \ ;\ \ a_3=a_3$

\begin{equation}
a_1a_3=-k^2\ \ ;\ \ A_0-\frac{k}{a_1}A_1=0
\label{4.1}\end{equation}

\item R-symmetry $a_1=a_3\ \ ;\ \ a_2=a_2$

\begin{equation}
a_1=0\ \ ;\ \ a_2=k\ \ ;\ \  A_0=0\ \ ;\ \  A_1\neq 0 
\label{4.2}\end{equation}
\begin{equation}
a_1=0\ \ ;\ \ a_2=-k\ \ ;\ \  A_0\neq 0\ \ ;\ \ A_1= 0 
\label{4.3}\end{equation}
\begin{equation}
a_1\neq 0\ ;\  a_2\neq\pm k\  ;\  a_1^2-a_2^2+k^2=0\  ;\  A_0-\frac{k-a_2}{a_1}A_1=0
\label{4.4}\end{equation}

\item generic solution (no discrete symmetry imposed)

\begin{equation}
a_1=\mbox{any} \ \ ;\ \ a_2=k\ \ ;\ \ a_3=0\ \ ;\ \ A_0=0\ \ ;\ \ A_1\neq 0 
\label{4.5}\end{equation}

\begin{equation}
a_1=0 \ \ ;\ \ a_2=k\ \ ;\ \ a_3\neq 0\ \ ;\ \ A_0+\frac{a_3}{2k}A_1=0
\label{4.6}\end{equation}

\begin{equation}
a_1=0 \ \ ;\ \ a_2=-k\ \ ;\ \ a_3=\mbox{any}\ \ ;\ \ A_0\neq 0\ \ ;\ \ A_1= 0 
\label{4.7}\end{equation}

\begin{equation}
a_1\neq 0 \ \ ;\ \ a_2=-k\ \ ;\ \ a_3=\mbox{any}\ \ ;\ \ A_0-\frac{2k}{a_1}A_1=0
\label{4.8}\end{equation}

\begin{equation}
a_1\neq 0\ ;\  a_2\neq\pm k\  ;\  a_3\neq 0\ ;\ a_1a_3-a_2^2+k^2=0\  ;\  A_0-\frac{k-a_2}{a_1}A_1=0
\label{4.9}\end{equation}
\end{enumerate}
Summarizing, three distinct types of boundary conditions on the fields are found, each realized by means of a particular choice of the coefficients appearing in  $S_{bd}$ \eqref{2.9}: 
\begin{enumerate}
\item
\begin{equation}
\left.A_0\right|_{x_2=0}=0 \ \ ;\ \ \left.A_1\right|_{x_2=0}\neq 0 
\label{4.10}\end{equation}
\item 
\begin{equation}
\left.A_0\right|_{x_2=0}\neq 0  \ \ ;\ \ \left.A_1\right|_{x_2=0}=0 
\label{4.11}\end{equation}
\item
\begin{equation}
\left.A_0-v A_1\right|_{x_2=0} = 0 \ ,
\label{4.12}\end{equation}
\end{enumerate}
where the coefficient $v$ in the third type of boundary condition \eqref{4.12}  depends on the details of the corresponding $S_{bd}$ \eqref{2.9}. In particular 
$v=\frac{k}{a_1}$ for the solution \eqref{4.1}, 
$v=\frac{k-a_2}{a_1}$ for  \eqref{4.4} and \eqref{4.9}, 
$v=-\frac{a_3}{2k}$ for  \eqref{4.6} and
$v=\frac{2k}{a_1}$ for  \eqref{4.8}.

\section{2D algebra and boundary action}

Deriving the Ward identity \eqref{3.4} with respect to $J_k(x')$ and going on-shell, $i.e.$ putting $J=0$, we have
\begin{eqnarray}
\partial_k\delta^{(2)}(X-X') &=& 
-k\epsilon_{ij}\partial_i\left\langle A_j(X)A_k(X')\right\rangle\nonumber\\
&=&-k\delta(t-t')\left[A_1(X),A_k(X')\right]\nonumber\\
&&-k\left\langle\epsilon_{ij}\partial_iA_j(X)A_k(X')\right\rangle \ ,
\label{5.1}\end{eqnarray}
where $X=X_i=(x_0,x_1)$ denote the 2D coordinates on the boundary $x_2=0$. From the Ward identity \eqref{3.4}, going on-shell, we also get 
\begin{equation}
\left. \epsilon_{ij}\partial_iA_j(x)\right|_{x_2=0}=0,
\label{5.2}\end{equation}
so that the second term on the rhs of \eqref{5.1} vanishes and we are left with the Ka\c{c}-Moody algebra
\begin{equation}
\left[A_1(X),A_1(X')\right]\delta(t-t')  = -\frac{1}{k}\partial_1\delta^{(2)}(X-X'),
\label{5.3}\end{equation}
with central charge $c=\frac{1}{k}$.

Moreover, the most general solution of the condition \eqref{5.2} 
\begin{equation}
\left.A_i\right|_{x_2=0}=\partial_i\Phi(X)
\label{5.4}\end{equation}
defines the scalar field $\Phi(X)$ as the dynamical variable which lives on the boundary $x_2=0$ of the bulk CS action $S_{bulk}$ \eqref{2.2}.

In terms of the scalar field $\Phi(X)$, the Ka\c{c}-Moody algebra \eqref{5.3} gives the boundary equal time commutators 
\begin{equation}
\delta(t-t')\left[-k\partial_1\Phi(X),\Phi(X')\right]=\delta^{(2)}(X-X'),
\label{5.5}\end{equation}
and, by comparison with the canonical equal time commutators
\begin{equation}
[q(X),p(X')]=\delta^{(2)}(X-X')
\label{5.6}\end{equation}
we are led to identify
\begin{equation}
q(X)\equiv -k\partial_1\Phi(X)\ \ ;\ \ p(X)\equiv\Phi(X).
\label{5.7}\end{equation}
It easy at this point to write the dynamical part of the 2D action induced by the canonical field variables \eqref{5.7}
\begin{equation}
S=\int d^2X\ {\cal L}=\int d^2X\  p\dot{q}=\int d^2X\  \Phi(X)\partial_0\left(  -k\partial_1\Phi(X)\right).
\label{5.8}\end{equation}

In order to find the more general 2D boundary action $S_{2D}$ we must take into account the following features

\begin{enumerate}

\item
$S_{2D}$ must depend on the 2D scalar potential $\Phi(X)$.

\begin{equation}
S_{2D} = S_{2D} [\Phi(X)],
\label{5.9}\end{equation}
whose canonical mass dimension   is zero $[\Phi]=0$.

\item

$S_{2D}$ must preserve the canonical commutation relations \eqref{5.5}. Hence it must not contain time derivative else than those appearing in \eqref{5.8}.

\item
The definition \eqref{5.4} of the 2D scalar potential $\Phi(X)$ is left invariant by the shift invariance
\begin{equation}
\delta_{shift}\Phi(X)=\eta\ ,
\label{5.10}\end{equation}
where $\eta$ is a constant. Therefore, we ask that the 2D boundary action is invariant under the shift symmetry \eqref{5.10} as well

\begin{equation}
\delta_{shift}S_{2D}=0\ .
\label{5.11}\end{equation}

\item
The equation of motion derived from $S_{2D}$ must be compatible with the boundary conditions \eqref{4.10}, \eqref{4.11} and \eqref{4.12}, which, written in terms of the 2D scalar potential $\Phi(x)$ become, respectively
\begin{enumerate}
\item
\begin{equation}
\partial_0\Phi=0\ \ ;\ \ \partial_1\Phi\neq 0
\label{5.12}\end{equation}
\item
\begin{equation}
\partial_0\Phi\neq0\ \ ;\ \ \partial_1\Phi= 0
\label{5.13}\end{equation}
\item
\begin{equation}
\partial_0\Phi -v \partial_1\Phi=0.
\label{5.14}\end{equation}
\end{enumerate}
\end{enumerate}
The most general $S_{2D} $ satisfying the constraints \eqref{5.9}, \eqref{5.8} and \eqref{5.11} is
\begin{equation}
S_{2D} = \int d^2X\ \left( -k\Phi\partial_0\partial_1\Phi - \frac{c}{2}(\partial_1\Phi)^2\right),
\label{5.15}\end{equation}
where $c$ is a constant which must be related to the details of the CS boundary action $S_{bd}$ \eqref{2.9} by asking that the equations of motion of the 2D scalar field $\Phi(X)$ are compatible with the boundary conditions \eqref{5.12}, \eqref{5.13} or \eqref{5.14}.

The equation of motion for the 2D scalar potential $\Phi(X)$ is
\begin{equation}
\frac{\delta S_{2D}}{\delta\Phi} = 
\partial_1(-2k\partial_0\Phi + c \partial_1\Phi)=0
\label{5.16}\end{equation}

which is compatible only with the class of boundary conditions \eqref{5.14}, if
\begin{equation}
c=2kv
\label{5.17}\end{equation}
which relates the 3D CS action with boundary $S_{tot}$ \eqref{2.6} with the 2D action $S_{2D}$ \eqref{5.15} (remember that $v=v(k;a_1,a_2,a_3)$).

Let's take for instance the solution \eqref{4.1} $a_1a_3=-k^2$, $a_2=0$, $A_1=\frac{a_1}{k}A_0\Rightarrow v=\frac{k}{a_1}$, which corresponds to imposing the P-symmetry on the boundary action $S_{bd}$ \eqref{2.9}. Compatibility between the 2D equation of motion \eqref{5.16} and the 3D boundary condition \eqref{5.14} is obtained if
\begin{equation}
c = \frac{2k^2}{a_1}.
\label{5.18}\end{equation}

The 2D action induced by the CS theory with a non Lorentz but parity invariant boundary is then
\begin{equation}
S_{2D}=k\int d^2X\ \left(\partial_0\Phi-\frac{k}{a_1}\partial_1\Phi\right)\partial_1\Phi\ .
\label{5.19}\end{equation}

The boundary conditions $A_i=0\rightarrow\partial_i\Phi=0$ \eqref{5.12} and  \eqref{5.13} are never compatible with a 2D action $S_{2D}$. The only acceptable boundary condition resulting from putting a boundary on the 3D CS action is that of the type \eqref{4.12} or, equivalently, \eqref{5.14}: $A_0-v A_1=0\rightarrow \partial_0\Phi-v\partial_1\Phi=0$.

\section{Contact with the fermion-boson correspondence and the Tomonaga-Luttinger theory}

It is has been already observed that the boundary conditions derived by imposing a boundary in TQFT represent a kind of duality relations between bosons and fermions \cite{Amoretti:2013nv} which leads to study the dynamics of dimensionally reduced topological models \cite{Amoretti:2012kb}. Summarizing, the presence of duality relations of the type \eqref{5.14} signalize that  the degrees of freedom living on a boundary of a bosonic TQFT are, indeed, fermionic \cite{Cho:2010rk}. Such relations have even been used to explicitly recover the fermionic degrees of freedom from the bosonic ones \cite{Aratyn:1984jz,Aratyn:1983bg,Amoretti:2013xya}.

It is quite remarkable that, because of the boundary/chirality/duality condition \eqref{5.14}, the scalar bosonic degree of freedom found on the 2D boundary represents indeed a chiral Weyl fermion.

Given a Weyl fermion $\psi(x,t)=\psi(X)$ in 1+1 dimensions, the normal ordered expression for the density operator is
\begin{equation}
\rho(x)=:\psi^\dagger(x)\psi(x):\ ,
\label{6.1}\end{equation}
which satisfies the commutation relation
\begin{equation}
[\rho(x),\rho(x')]=\frac{\nu}{4\pi}\delta'(x-x'),
\label{6.2}\end{equation}
where $\nu$ is the filling factor of the Fractional Quantum Hall Effect (FQHE). The relation \eqref{6.2}, as it is well known, characterizes the Tomonaga-Luttinger theory for a 1+1 dimensional liquid of interacting electrons \cite{Tomonaga:1950zz,Luttinger:1963zz,Haldane:1981zza}. All local operators ${\cal O}(x,t)$ of the theory of the Weyl fermion $\psi(x,t)$ are chiral, $i.e.$ satisfy
\begin{equation}
{\cal O}(x,t)={\cal O}(x-vt).
\label{6.3}\end{equation}
The fermion-boson correspondence is based on the fact that the operator density $\rho(x,t)$ can be written in terms of a  chiral boson $\Phi(x,t)=\Phi(x-vt)$
\begin{equation}
\rho(x,t)=\frac{1}{2\pi}\partial_x\Phi(x,t),
\label{6.4}\end{equation}
whose corresponding bosonic Hamiltonian is 
\begin{equation}
H=\frac{v}{4\pi\nu}\int dxdt_1(\partial_1\Phi)^2\ ,
\label{6.5}\end{equation}
which indeed describes the edge states in the FQHE for the Laughlin's sequence for the filling factor $\nu$ \cite{Wen:1990se}.

All these features are naturally recovered in the framework of a topological field theory with boundary. From the 2D action \eqref{5.15}, where the constant $c$ is given by \eqref{5.17}, we immediately get the 2D Hamiltonian
\begin{equation}
H=kv\int d^2X(\partial_1\Phi)^2\ ,
\label{6.5.1}\end{equation}
which coincides with the FQHE Hamiltonian \eqref{6.5} provided that the CS coupling $k$ and the filling factor $\nu$ are related as follows
\begin{equation}
k=\frac{1}{4\pi\nu}\ .
\label{6.5.2}\end{equation}
The formula for the current anomaly \eqref{6.2} coincides with the Ka\c{c}-Moody algebra \eqref{5.3} where the CS boundary field $A_1(X)$ $is$ the density operator $\rho(x,t)$, and the solution of the boundary condition \eqref{5.4} coincides with the bosonization relation \eqref{6.4}. The chirality of the boson is ensured by the boundary condition \eqref{5.14}, which is compatible with the equation of motion of the bosonic 2D bosonic action \eqref{5.16}, which directly states the chirality of the density operator \eqref{6.1}. Besides this, on the condensed matter counterpart, the equation of motion for the 2D scalar potential \eqref{5.16} is the continuity equation for the Tomonaga-Luttinger liquid. In fact, the equation of motion \eqref{5.16}, written in terms of the density operator $\rho(x,t)$ \eqref{6.4}, reads
\begin{equation}
\partial_0\rho+\partial_1(-\frac{c}{2k}\rho)=0\ ,
\label{6.6}\end{equation}
which, by comparison with
\begin{equation}
\partial_0\rho+\partial_1J=0\ ,
\label{6.7}\end{equation}
leads to identify the current
\begin{equation}
J\equiv -v\rho\ ,
\label{6.}\end{equation}

\section{Conclusions}

In this paper we added a planar, single-sided boundary to the 3D CS theory, without computing the propagators, which is an essential task in the Symanzik method \cite{Symanzik:1981wd} used previously \cite{Blasi:2008gt}, which could be unfeasible in more general cases. 

A central role is played by the boundary term of the 3D action \eqref{2.9}. It turns out that it is necessary to break Lorentz invariance even on the boundary plane, and this distinguishes our approach from the usual one \cite{Amoretti:2014iza}. In fact, if Lorentz invariance on the 2D boundary is imposed, no compatible boundary conditions for the fields can be obtained. The method we used has been inspired by the procedure followed in \cite{Rubakov:2004eb,Libanov:2005vu} in the quite different framework of massive gravity. Two discrete symmetries -``parity'' and ``time-reversal'' - of the whole theory (3D bulk with 2D boundary) have been found. Nontrivial results are found by imposing either one or none of them. In fact, asking the presence of both discrete symmetries is equivalent to asking 2D Lorentz invariance on the boundary, and no solution exists in that case. This kind of decomposition of Lorentz invariance in terms of two discrete symmetries seems to be peculiar to the boundary of the CS theory. 

In \cite{Blasi:2008gt} the physics on the boundary has been cleverly inferred by exploiting the analogy between the algebra of chiral currents found on the boundary and the bosonic chiral propagator. In this paper, using symmetry arguments, it has been found that the degree of freedom on the boundary is a scalar \eqref{5.4}, obeying necessarily the boundary condition \eqref{5.14}, which, in turn, has been derived from the equations of motion of the 3D CS action with boundary. 

The Ka\c{c}-Moody current algebra \eqref{5.3} has been used to identify the canonical conjugate field variables by means of which the 2D action for the scalar field \eqref{5.15} has been constructed, using also the shift symmetry \eqref{5.11} which emerges naturally from our construction. The corresponding 2D Hamiltonian coincides with the one proposed by Wen \cite{Wen:1990se} for the edge states of the FQHE, and the relation \eqref{6.5.2} between the CS coupling and the filling factor has been derived, without being inferred by analogy as done in \cite{Blasi:2008gt}. 

The unique boundary condition \eqref{5.14} plays several roles in the scenario described in this paper. Firstly, it represents a chirality condition for the edge states. This corresponds to the fact that the edge FQHE currents follow preferred directions. Secondly, the boundary condition is a bosonization condition, which tells us that the scalar boson we found on the boundary is, indeed, a Weyl chiral fermion. Finally, we found that that same boundary condition  is also the continuity equation involving a current density and a charge density, which we were able to identify in terms of the scalar degree of freedom living on the 2D boundary of the 3D CS theory,  making the link between this latter model and the Luttinger-Tomonaga theory  once more evident.

{\bf Acknowledgements}

N.M. thanks the support of INFN Scientific Initiative SFT: ``Statistical Field Theory, Low-Dimensional Systems, Integrable Models and Applications''


\end{document}